\documentclass[10pt, conference, compsocconf]{IEEEtran}
\IEEEoverridecommandlockouts
\usepackage{cite}
\usepackage{amsmath,amssymb,amsfonts}
\usepackage{algorithmic}
\usepackage{graphicx}
\usepackage{textcomp}
\usepackage{xcolor}
\usepackage{tikz}
\usepackage{pgfplots}
\usepackage{verbatim}
\usepackage{dblfloatfix}
\usepackage{multirow}
\usepackage{url}
\usepackage{booktabs}
\usepackage{eucal}

\newcommand\preserveCal[2]{
    \expandafter\newsavebox\csname box@\string#1\endcsname
    \expandafter\savebox\csname box@\string#1\endcsname{\ensuremath{\mathcal{#2}}}
    \expandafter\def\expandafter#1\expandafter
        {\expandafter\usebox\csname box@\string#1\endcsname}
}
\preserveCal{\calD}{D}
\preserveCal{\calG}{G}
\preserveCal{\calI}{I}
\preserveCal{\calP}{P}
\preserveCal{\calU}{U}
\preserveCal{\calV}{V}

\makeatother

\def\BibTeX{{\rm B\kern-.05em{\sc i\kern-.025em b}\kern-.08em
    T\kern-.1667em\lower.7ex\hbox{E}\kern-.125emX}}

\begin{document}

\title{Parallel/distributed implementation of cellular training for generative adversarial neural networks
}

\author{\IEEEauthorblockN{Emiliano P\'erez}
\IEEEauthorblockA{\textit{Universidad de la Rep\'ublica}\\ Uruguay \\
emiliano.perez@fing.edu.uy}
\and
\IEEEauthorblockN{Sergio Nesmachnow}
\IEEEauthorblockA{\textit{Universidad de la Rep\'ublica}\\ Uruguay \\
sergion@fing.edu.uy}
\and
\IEEEauthorblockN{Jamal Toutouh, Erik Hemberg, Una-May O'Reilly}
\IEEEauthorblockA{\textit{Massachusetts Institute of Technology}\\
Cambridge, MA, USA \\
toutouh@mit.edu,\{hembergerik,unamay\}@csail.mit.edu}
}


\maketitle


\begin{abstract}
Generative adversarial networks (GANs) are widely used to learn generative models. 
GANs consist of two networks, a generator and a discriminator, that apply adversarial learning to optimize their parameters. This article presents a parallel/distributed implementation of a cellular competitive coevolutionary method to train two populations of GANs. A distributed memory parallel implementation is proposed for execution in high performance/supercomputing centers. Efficient results are reported on addressing the generation of handwritten digits (MNIST dataset samples). Moreover, the proposed implementation is able to reduce the training times and scale properly when considering different grid sizes for training.
\end{abstract}


\noindent\fbox{%
    \parbox{\linewidth}{%
    \footnotesize
        \textbf{Copyright notice:} This article has been accepted for publication in IEEE  International Parallel and Distributed Processing Symposium, Parallel and Distributed Combinatorics and Optimization, 2020.}%
}

\normalsize

\section{Introduction}
Generative machine learning has been demonstrated being a successful tool for a wide range of applications~\cite{wu2017survey,pan2019recent}.
Generative Adversarial Networks (GANs) is a powerful method for such type of machine learning~\cite{goodfellow2014generative}.
GANs take a training set drawn from a specific distribution and learn to represent an estimate of that distribution. 
In general, they consist of two neural networks, a generator and a discriminator, that applies adversarial learning to optimize their parameters. 
The discriminator learns how to distinguish the ``natural/real'' samples from the ``artificial/fake'' samples produced by the generator. 
The generator is trained to transform its inputs from a random latent space into ``artificial/fake'' samples to fool the discriminator. 
GAN training is formulated as a minmax optimization problem by the definitions of generator and discriminator loss~\cite{goodfellow2014generative}.

The GAN training can converge on a generator that is able to approximate the real distribution so well that the discriminator only provides a random label for real and fake samples. 
Nevertheless, training GANs is difficult since the adversarial dynamics may give rise to different convergence pathologies~\cite{arjovsky2017towards}, e.g., gradient explosion and mode collapse.
When gradient pathologies appear, the generator is not able to learn and, if the problem persists, it basically generates noise for the whole training process. 
Mode (generator) collapse happens when the training converges to a local optimum, 
i.e. the generator produces realistic fake images that only represent a portion of the real data distribution. Therefore, the GAN has not successfully learned the distribution. 

Distributed coevolutionary algorithms have shown to be effective overcoming GAN training pathologies. 
They train two populations, one of generators and one of discriminators, which are trained by fostering an arm-race between them. 
One of the main issues of such a methods is the scalability since they train populations of networks which requires high computational costs. 
In order to address this problem, Lipizzaner
\cite{schmiedlechner2018lipizzaner} 
and Mustangs~\cite{toutouh2019} apply a spatially distributed coevolutionary algorithm to reduce the number of fitness evaluations that competitive coevolution may require. 

Nowadays, a common infrastructure for scientific computing is provided by large high performance/supercomputing centers, which gather a significantly large number of computing resources 
dedicated to research and development~\cite{Foster1995}. These infrastructures allow sharing resources between the academic community, providing a cost-effective and rational utilization of scarce monetary funding for research~\cite{Nesmachnow2019}

In this line of work, this article presents a parallel/distributed implementation of cellular training for GANs. The proposed implementation is based on a distributed memory approach to be executed on (non-dedicated) high performance/supercomputing centers. A two-level parallel model is applied, using 
multithreading programming and distributed memory computing via the Message Passing Interface (MPI).

Thus, the main contributions of the research reported in this article are: i) a distributed memory parallel implementation of the Mustangs/Lipizzaner framework for GANs training, ii) the experimental evaluation for a relevant case study for GANs, MNIST dataset samples generation~\cite{lecun1998mnist}, i.e., the widely used generation of handwritten digits from zero to nine.

The article is organized as follows. Section~\ref{Sec:GAN-training} introduces the problem of GANs training
and a brief review of related works.
The proposed parallel implementation for GANs training is described in Section~\ref{Sec:Implementation}. Section~\ref{sec:experiments} reports the experimental evaluation of the proposed implementation for Mustangs/Lipizzaner. Finally, Section~\ref{Sec:Conc} presents the conclusions and the main lines for future work.

\section{Distributed Coevolutionary GANs Training}
\label{Sec:GAN-training}

Robust GAN training is still an open research
question~\cite{arora2017gans}.
This section introduces the optimization problem addressed to train 
GAN and the spatial model for training.

\subsection{General GAN training}
\label{sec:gan-training}
Let $\calG$=$\{G_g, g \in \calU\}$ and $\calD$=$\{D_d, d \in \calV\}$ denote the class of generators and discriminators, where $G_g$ and $D_d$ are functions
parameterized by $g$ and $d$. $\calU, \calV \subseteq R^{p}$
are
the 
parameters space of the generators and
discriminators. Finally, let $G_*$ be the target unknown distribution to 
fit the 
generative model
\cite{arora2017gans}

Formally, the goal of GAN training is to find parameters $g$ and $d$ to optimize the objective function $\min_{g\in \calU}\max_{d \in \calV} \mathcal{L}(g,d)$, where $\mathcal{L}(g,d)$ = $E_{x\sim G_*}[\phi (D_d(x))] + E_{x\sim G_g}[\phi(1-D_d(x))]$ and $\phi$:[0,1] $\to R$, is a concave \emph{measuring function}. In practice, a finite number of training samples $x_1, \ldots, x_m \sim G_*$ is available. Therefore, an empirical version $\frac{1}{m}\sum_{i=1}^{m} \phi(D_d(x_i))$ is used to estimate $E_{x\sim G_*}[\phi (D_d(x))]$. The same also holds for $G_g$.

\subsection{Spatial GAN Training}
\label{sec:spatial-training}

Mustangs/Lipizzaner~\cite{schmiedlechner2018towards,toutouh2019} adversarially trains a population of generators $\mathbf{g}$=$\{g_1, ..., g_N\}$ and a population of discriminators $\mathbf{d}$=$\{d_1, ..., d_N\}$, one against the other ($N$ is the size of the population). The idea is to apply the benefits of population based competitive coevolution to address GAN training pathologies. 

\begin{figure}[!h]
\setlength{\abovecaptionskip}{3pt}
  \includegraphics[width=1\linewidth]{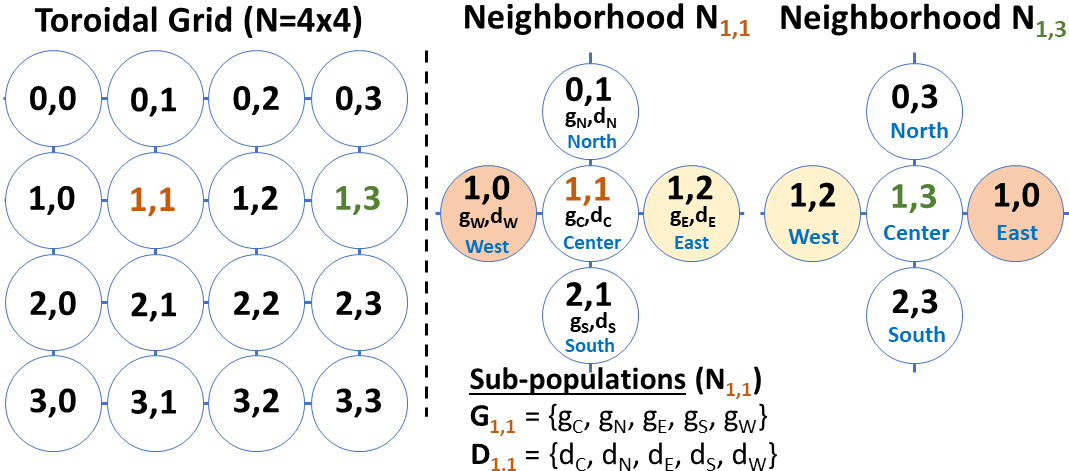}
  \caption{Left: 4$\times$4 grid; right:  neighborhoods N$_{1,3}$, N$_{1,1}$ with sub-populations.}
  \label{fig:4x4-nhood}
\end{figure}

Distributed training defines a toroidal grid in whose cells a GAN is placed (called \textit{center}). This allows defining neighborhoods with sub-populations of generators 
and discriminators 
to mitigate the quadratic computational complexity of the adversarial paradigm applied. 
The size of the sub-populations is denoted by $s$ ($s \leq N$). Without loss of generality, square grids of $m$ $\times$ $m$ are considered, i.e., there are $m^2$ neighborhoods. In this article, a five-cell Moore neighborhood ($s$=5) is used, i.e., the neighborhoods include the cell itself (center) and the cells in the {West}, {North}, {East}, and {South} (see Fig.~\ref{fig:4x4-nhood}).

Each cell executes its own learning algorithm 
by applying an asynchronous parallel training to optimize the 
sub-populations of generators and discriminators. Cells interact with neighbors after each training epoch to exchange the {center} GAN among neighborhoods to update the sub-populations. Communications are carried out through overlapping neighborhoods in the spatial grid. Thus, each neighborhood gathers the latest updated {center} generator and discriminator of its overlapping neighborhoods. 
Fig.~\ref{fig:4x4-nhood} illustrates examples of the overlapping neighborhoods on a 4$\times$4 toroidal grid. Updates in cell N$_{1,0}$ and N$_{1,2}$ will be communicated to the neighborhoods in the range of such a cell, e.g., N$_{1,1}$ and N$_{1,3}$. 

At the end of the method, the generative model returned is the one defined by the sub-population with the highest quality according to some fitness value, e.g., inception score. 

\subsection{Related Work}
\label{Sec:RW}

Robust GANs training is still an open research
topic
\cite{arora2017gans}.
This special type of adversarial learning frequently shows problems or pathologies~\cite{arora2017generalization,arjovsky2017towards}, 
e.g., {mode collapse}, {discriminator collapse}, and {vanishing gradients}. 
The main reason is that optimizing the {minmax GAN objective} is generally performed by simultaneous gradient-based updates to the parameters of the networks that hardly converges to an equilibrium.
Theoretical models have been proposed to provide a better understanding of dynamics when training generators against discriminators and vice-versa.

The use of multiple generators and/or discriminators for improving training robustness has also been studied. Some of the proposals include training a cascade of GANs~\cite{adlam2019learning},
sequentially training and adding new generators with boosting
techniques~\cite{tolstikhin2017adagan}, training multiple generators
and discriminators in parallel~\cite{schmiedlechner2018towards,toutouh2019}, 
training
an array of discriminators specialized in a different low-dimensional
projection of the data~\cite{neyshabur2017stabilizing}, and 
using several
adversarial ``local'' pairs of networks that are trained independently so
that a ``global'' supervising pair of networks can be trained against
them~\cite{chavdarova2018sgan}.

\section{Parallel/distributed \\implementation of GAN training}
\label{Sec:Implementation}

This section describes the proposed parallel/distributed implementation of cellular training for GANs.

\subsection{Parallel model}

An adaptation of the traditional master-slave model for parallel computing is applied for the proposed implementation of Mustangs/Lipizzaner. This model provides a simple, yet effective, concurrent processing model to achieve good scalability to take advantage of large parallel computing environments. Furthermore, the model is highly flexible and adaptable, allowing to develop multi-level parallel implementations using both shared-memory and distributed-memory approaches~\cite{Foster1995}. 

In the proposed parallel model,
the master process controls a group of slaves. Domain decomposition 
considers
the grid using for 
GAN training.
A uniform partitioning criteria is applied, since the estimated workload (thus, also the expected execution time) 
within each 
cell
is the same. In case that a different criteria is used (i.e., for taking advantage of heterogeneous computing platforms) a dynamic partitioning criteria is needed.
The proposed data-parallel approach associates the grid coordinates with the rank of each slave process in the context of the global \texttt{MPI\_COMM\_WORLD} communicator. Additional features, e.g. introducing a 
Cartesian topology via \texttt{\small MPI\_CART\_CREATE}, can be applied to 
optimize communications.
The main details of the proposed model 
are provided in the following subsections.

\subsection{Master and slave processes}

Two main processes are defined in the proposed parallel model. Both of them are implemented using multithread programming
to achieve the best performance and scalability.

The master process is in charge of controlling the execution flow of the program. The master performs several management and coordination tasks when the system starts execution, including: i) gathering the information about the computing infrastructure and computing resources available, ii) deciding in which node each slave process will execute, iii) assigning workload to each slave, applying a strategy oriented to minimize and balance the load on each node, and iv) sharing the parameter configuration to be used in the execution with all slave processes. After that, the master launches the slaves and starts their execution. During the execution, the master periodically performs control activities to determine if all slaves are working properly, are on time, or are delayed in its execution. This task is performed via a specific protocol using heartbeats 
(see next subsection). 
Heartbeats are handled by a 
thread of the master process (the \textit{heartbeat thread}), in order to perform the system monitoring in background, without interfering with the main processing of the system.
Finally, other tasks are performed by the master process once slaves end execution: gathering the processed local results from each slave, processing the 
intermediate results in a reduction phase, and returning the best result obtained overall.

In turn, each slave process is in charge of performing the 
GAN training.
In a previous 
implementation of Mustangs/Lipizzaner, each slave is binded to a port,
allowing the system to execute in a client-server parallel model. Instead, in the proposed implementation all the slaves 
assigned to work on specific training tasks join two communication channels (see next subsection) for communication with the master.

A slave waits for an order to start, from the master. After that, it receives the parameter configuration and starts execution. Training is performed considering the neighborhood information, 
accessed via the 
\texttt{\small grid} 
class and communicated in the 
parameter configuration. The first action performed by the slave is to launch a secondary execution thread to train the GAN, building a grid with the information of neighboring slaves. In turn, the main thread of execution is used as an interface for communications with the master. This interface allows the master to retrieve useful information from the execution (the \textit{status)} and it also allows performing other communications from/to the master (see next subsection).
Communications with the master are handled by the main thread of each slave process, while the training is performed by the \textit{execution thread}, 
in order to achieve concurrency.

\begin{figure}[!h]
\centerline{\includegraphics[scale=0.45]{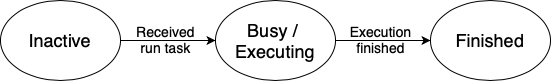}}
\caption{States and transitions of slave processes in the proposed parallel implementation of Mustangs/Lipizzaner}
\label{Fig:states}
\end{figure}

Slaves have three states: 
i) \textit{inactive},
the state in which a slave 
has not received a workload to process yet; ii) \textit{processing}, 
the state in which the slave is performing the assigned training; and iii) \textit{finished}, after the slave finish execution and is waiting for the master to gather the results.
%
%
A slave changes its state from inactive to processing after receiving a \textit{run task} message message from the master, In turn, a slave changes its state from processing to finished after performing the last 
iteration of the training process.
A diagram of states and transitions for slave processes is presented in Fig.~\ref{Fig:states}.

\setlength{\abovecaptionskip}{3pt}
\setlength{\belowcaptionskip}{3pt}

\subsection{New classes in the parallel implementation}

The proposed implementation has two main new classes:

\begin{itemize}
    \item Class \texttt{\small comm-manager} substitutes the original class \texttt{\small node-comm} of Mustangs/Lipizzaner. 
    Overall, \texttt{\small comm-manager} is a wrapper of all functions used for communication between processes, defined in an abstract way without defining explicitly how the communications are implemented.
    \item Class \texttt{\small grid} substitutes the original class \texttt{\small neighbourhood} of Mustangs/Lipizzaner. Grid is in charge of defining the grid for the execution of each slave process. A specific feature of the \texttt{\small grid} class is that it allows modifying the grid and also the structure of neighboring processes dynamically, a feature that was not provided by the original implementation of Mustangs/Lipizzaner. Dynamically changing the neighborhood allows exploring different patterns for training and learning. In addition,  class \texttt{\small grid} does not depend on \texttt{\small comm-manager}. The implementation is decoupled, so different modules for communication can be applied.
\end{itemize}

\begin{figure*}[b]
\setlength{\abovecaptionskip}{-3pt}
\setlength{\belowcaptionskip}{-3pt}
\centerline{\includegraphics[scale=0.75,trim=0cm 7cm 0cm 2.5cm, clip]{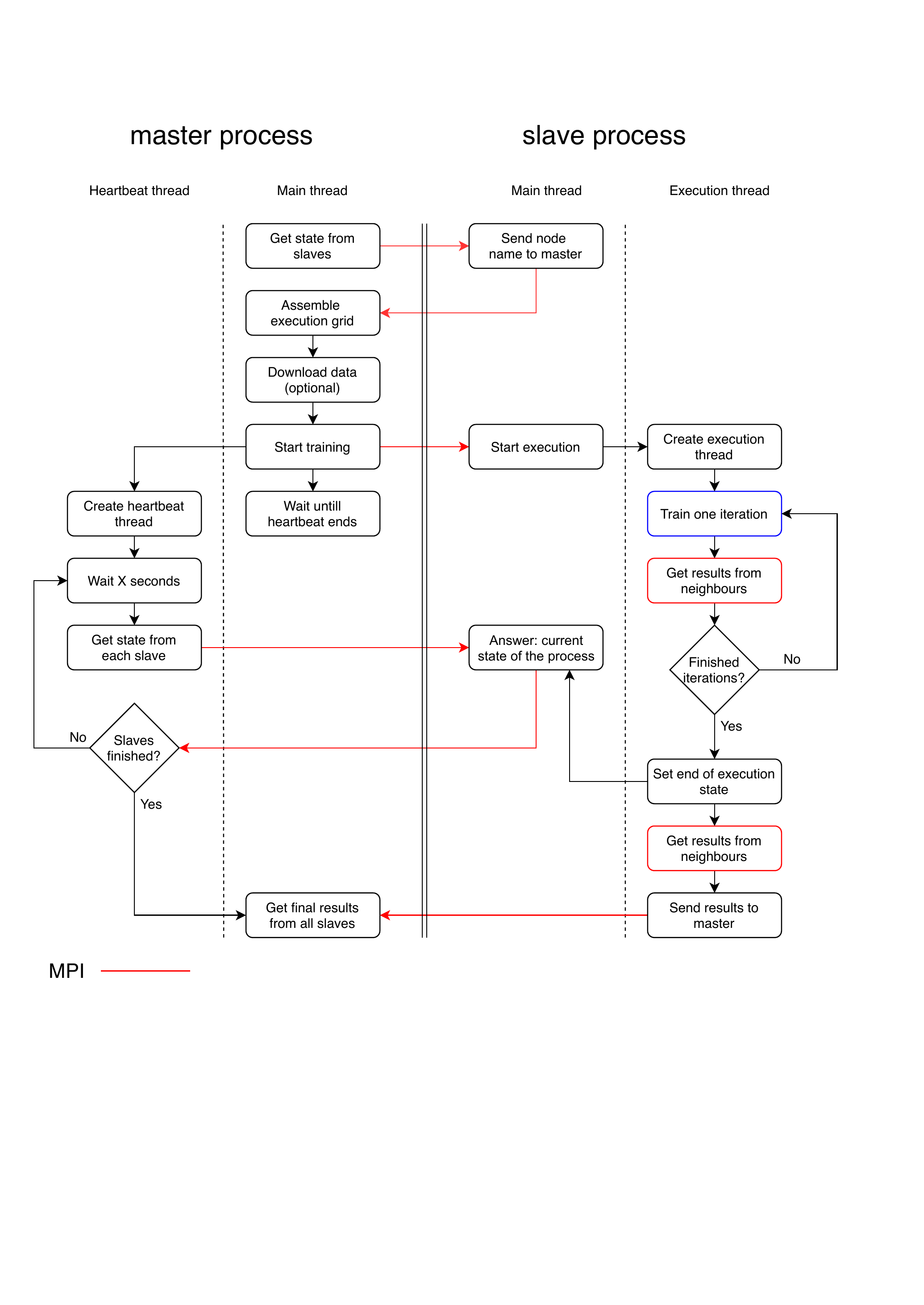}}
\caption{Flow diagram (processing and communications) between the master process and a representative slave process in the proposed implementation}
\label{Fig:flow-execution}
\end{figure*}

\subsection{Communications and synchronizations}

The proposed implementation was developed using Message Passing Interface (MPI)~\cite{Gropp2014}, allowing an easy deployment and instalation in nowadays large high performance computing platforms. The communication modes in MPI provide versatility for implementing different types of information exchange, for both sending messages and exchanging control and synchronization information.

Class \texttt{\small comm-manager} implements all the features and functions originally provided by class \texttt{\small node-comm}, but using the MPI library. \texttt{\small Comm-manager} implements all communications and synchronization in an abstract way, using underlying MPI functions for sending and receiving information between processing units.
Three contexts of cmmunications (i.e., MPI communicators) are used in \texttt{\small comm-manager}: 

\begin{itemize}
    \item The global \texttt{\small WORLD} communicator, which is used at the beggining of the system execution to establish all global configurations. 
    In addition, the \texttt{\small WORLD} communicator is used to exchange messages to start the execution of slaves (\textit{run task} messages) and also \textit{get status} control messages.
    \texttt{\small WORLD} is the base communicator to define all other communications contexts, too.
    \item The \texttt{\small LOCAL} communicator is defined for performing collective operations that only involve active slaves (in \textit{processing} state) in a given grid. 
    One of the main features of this communicator is allowing the execution of \texttt{\small gather} operations 
    without involving the master process or inactive slave processes. Gather operations are performed between slaves to collect partial results, since every slave need to know the training results from neighboring slaves 
    in each iteration
    to continue its processing.
    \item The \texttt{\small GLOBAL} communicator includes all slaves and the master process and is used to perform collective operations involving all processes, such as the master gathering the results obtained by each slave for further processing once all iterations are performed.
\end{itemize}


A flow diagram of execution (processing and communications) between the master process and a representative slave process is presented in Fig.~\ref{Fig:flow-execution}. Communications and synchronizations performed in MPI are marked in red lines. The reported implementation is designed to execute in CPU. However, the blue square in the training iteration represents a processing that can be performed in GPU, as proposed in one of the main lines for future work.

\section{Experimental evaluation}
\label{sec:experiments}

This section presents the experimental evaluation and the efficiency analysis for the sequential and parallel versions of the proposed algorithm.

\subsection{Methodology and parameters setting}

The experimental evaluation consisted on executing the developed parallel implementation of Mustangs/Lipizzaner for different grid sizes and analyzing the computational efficiency. The settings of the parallel implementation are described in Table~\ref{Table:settings}. The experiments evaluated the results using different grid sizes for the slaves to be allocated. 

\begin{table}[!h]
\setlength{\tabcolsep}{8pt}
\centering
\caption{Parameters settings of the trained GANs}    
\label{Table:settings}
\begin{tabular}{lr}
\toprule
\textit{parameter} & \textit{value} \\ 
\midrule
\multicolumn{2}{c}{\textit{Network topolog}y}  \\
\midrule
Network type    & MLP \\
Input neurons   & 64 \\
Number of hidden layers & 2 \\
Neurons per hidden layer& 256 \\
Output neurons  & 784 \\
Activation function & tanh \\ 
\midrule
\multicolumn{2}{c}{\textit{Coevolutionary settings}} \\ 
\midrule
Iterations  & 200 \\
Population size per cell  & 1 \\
Tournament size & 2 \\
Grid size   & 2$\times$2 to 4$\times$4 \\
Mixture mutation scale & 0.01 \\ 
\midrule
\multicolumn{2}{c}{\textit{Hyperparameter} mutation} \\ 
\midrule
Optimizer & Adam \\
Initial learning rate & 0.0002 \\
Mutation rate & 0.0001 \\
Mutation probability & 0.5 \\ \midrule
\multicolumn{2}{c}{\textit{Training settings}} \\ 
\midrule
Batch size & 100 \\
Skip N disc.~steps & 1 \\ 
\midrule
\multicolumn{2}{c}{Execution settings} \\ 
\midrule
Number of tasks & 5 to 17 \\
Time limit & 96 hours \\
Temporary storage & 40GB \\
\bottomrule
\end{tabular}
\end{table}

\subsection{Development and execution platform}

The proposed parallel/distributed implementation of Mustangs/Lipizzaner was implemented in Python3 using pytorch~\cite{Paszke2017}. The python interface mpi4py (\url{https://mpi4py.readthedocs.io/}) was used for the parallelization with MPI.

The experimental analysis was performed on the National Supercomputing Center (Cluster-UY), Uruguay~\cite{Nesmachnow2019}. Cluster-UY provides up to 30 computing servers, each of them with Xeon Gold 6138 processors with 40 cores, 128 GB of RAM memory and 300GB of SSD storage for temporary files. The platform uses slurm (\url{https://slurm.schedmd.com/}) to manage the resources allocated to each job. 

A summary of the resource allocation used for each experiment is reported on 
Table~\ref{Table:resources}.
In each experiment, one (master or slave) process executes in a unique core.
Cluster-UY is a collaborative high performance computing platform and a best-effort queue is used, thus the availability of computing resources on the same node is no guaranteed.
All values reported in this section correspond to grid sizes of 2$\times$2, 3$\times$3 and 4$\times$4. Ten executions were performed for each experiment, in order to reduce the effects of non-determinism in resource allocation and parallel execution. Average and standard deviation values are computed for the obtained execution times.

\begin{table}[!h]
\setlength{\tabcolsep}{12pt}
\caption{Summary of resources used on each execution in the experimental analysis}
\label{Table:resources}
\centering
\begin{tabular}{lrrr}
\toprule
\multirow{2}{*}{\textit{parameter}} & \multicolumn{3}{c}{\textit{grid size}}\\
\cline{2-4}\\[-7pt]
& 2$\times$2 & 3$\times$3 & 4$\times$4 \\ 
\midrule
\# cores & 5 & 10 & 17 \\
memory (MB) & 9216 & 18432 & 32768\\
\bottomrule
\end{tabular}
\end{table}

\subsection{Test problem and instances}
The proposed implementation was evaluated using the widely used MNIST dataset~\cite{lecun1998mnist}. It consists of low dimensional handwritten digits (from zero to nine) images. 
The dataset consists of 70,000 images (samples): 60,000 samples for the training set and 10,000 for the test set. 
The samples are grayscale images, size-normalized, and centered in a fixed-size of 28$\times$28.

This image dataset is used in GAN literature to asses the generative modeling performance since it is appropriate for investigating mode collapse due to its limited target space and gradient vanishing problems~\cite{toutouh2019,kos2018adversarial, Toutouh2020_book, liu2019design}. 

\subsection{Efficiency results}

The execution times obtained for the single core and the parallel/distributed version of Mustangs/Lipizzaner are reported in Table~\ref{Tab:execution-times}. Execution times for the parallel/distributed version correspond to average and standard deviation obtained in the ten independent execution performed for each grid size. All times are reported in minutes. In addition, the speedup, defined as the ratio between the execution times of the single core and the distributed implementation, is reported.

\begin{table}[!h]
\setlength{\tabcolsep}{8pt}
\caption{Summary of execution times of GAN training}
\label{Tab:execution-times}
\centering
\begin{tabular}{crrr}
\toprule
\textit{grid size} & \multicolumn{1}{c}{\textit{single core (min)}} & \multicolumn{1}{c}{\textit{distributed }} & 
\multicolumn{1}{c}{\textit{speedup}}\\
\midrule
2$\times$2 & 339.6  & 39.81$\pm$0.01  &  8.53  \\
3$\times$3 & 999.5  & 73.24$\pm$2.56  & 13.65 \\
4$\times$4 & 1920.0 & 126.68$\pm$3.42 & 15.17 \\ 
\bottomrule
\end{tabular}
\end{table}

Results in Table~\ref{Tab:execution-times} indicates that the proposed parallel/distributed implementation is able to significantly reduce the execution times of GAN training using Mustangs/Lipizzaner. Speedup values were up to \textbf{15.17} for the execution times using a grid of size 4$\times$4, demonstrating a very good scalability behavior of the proposed implementation.
Superlinear speedup values were obtained for problem instance dimensions (grid size) 2$\times$2, and 3$\times$3, mostly do efficient management of the required memory for training. When using more computing resources, speedup reduce to sublinear values, due to the overhead introduced by process management and the implemented communications between processes. Anyway, the speedup values obtained for a grid of size 4$\times$4 allow reducing the execution times from 1920 minutes to 129 minutes, significantly increasing the applicability of the methodology, especially when new training are needed.

Table~\ref{Tab:profiling-times} presents a summary of the results of the profiling performed to the developed implementation. The execution time for the four most time consuming routines detected in the profiling (mutation, updating genome information, training, and gathering information) for single core and the distributed version. Mutate, training, and update\_genome are 
functions of the GAN training as implemented in pytorch, while gather information (using MPI\_allgather routine) 
is specific of the
parallel implementation. Results correspond to the average execution times for a grid size of 4$\times$4, which is representative of the efficiency results obtained in all the experiments.
The acceleration and speedup for each routine are reported. Acceleration values indicate the reduction on the execution times with respect to the single core execution.

\begin{table}[!h]
\setlength{\belowcaptionskip}{0pt}
\setlength{\abovecaptionskip}{0pt}
\setlength{\tabcolsep}{5pt}
\caption{Profiling of execution times \break for the most consuming routines in GAN training (minutes)}
\label{Tab:profiling-times}
\centering
\begin{tabular}{crrrr}
\toprule
\textit{routine} & \textit{single core} & \textit{distributed} & \textit{acceleration} & \textit{speedup} \\
\midrule
gather & 19.4 & 19.4 & 0.0\% & 1.00   \\
train & 264.9 & 43.8 & 83.5\% & 6.05   \\
update\_genomes & 199.8 & 16.8 & 91.6\% & 11.87  \\
mutate & 25.6 & 17.9 & 29.9\% & 1.43   \\
\midrule
overall & 509.6 & 97.9 & 80.8\% & 5.21 \\
\bottomrule
\end{tabular}
\end{table}

\begin{figure}[!h]
\includegraphics[width=1\linewidth]{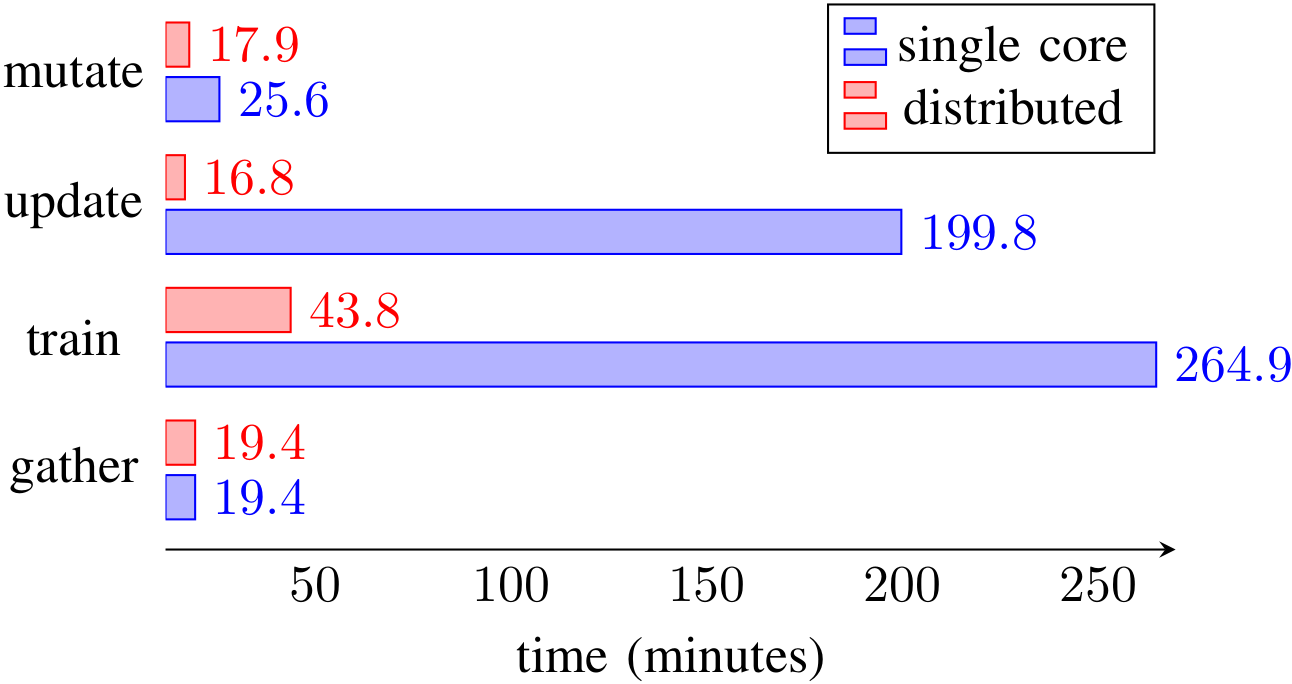}
\caption{Execution time comparison for single-node and parallel versions of the main routines in Mustang/Lipizzaner}
\label{Fig:profiling-times}
\end{figure}

Fig.~\ref{Fig:profiling-times} presents a graphical comparison of the main four routines in the processing.

Results reported in Table~\ref{Tab:profiling-times} and Fig.~\ref{Fig:profiling-times} clearly shows that the distributed implementation is able to significantly reduce the computing time of the most demanding routines (update and train). Speedup values of up to 6.05 and 11.87 were obtained for each routine, respectively. Furthermore, the time demanded for communications remained the same for both variants, suggesting a good scalability behavior of the proposed implementation.

\section{Conclusions}
\label{Sec:Conc}

This article presented a parallel distributed implementation of Mustangs/Lipizzaner, a framework for GAN training.
The proposed implementation uses the MPI library and multithreading programming for implementing a versatile and efficient version of Mustangs/Lipizzaner using both shared-memory and distributed-memory approaches, suitable for executing in nowadays high performance computing systems.

Results obtained in the experimental evaluation of the proposed parallel implementation indicate that it is able to effectively reduce the execution times of GANs training, while demonstrating a robust and efficient scalability behavior. Speedup values of up to 15.17 were obtained, for a training instance using a grid of size 4$\times$4. Furthermore, the profiling performed to the proposed implementation indicated that it reduces the execution time of the most demanding routines in the training.

The main lines for future work are related to improving the computational efficiency of the proposed implementation, taking into account detailed information from the profiling performed in distributed executions, and developing an efficient hybrid CPU/GPU version of the proposed method to take full advantage of nowadays scientific computing platforms. 
Moreover, we want to apply our method to train GANs to address de generation of higher dimensional images, such as samples from CIFAR and CelebA.

\small
\section*{Acknowledgment}
The work of S. Nesmachnow is partly supported by ANII and PEDECIBA, Uruguay. J.~Toutouh has been partially funded by EU Horizon 2020 research and innovation programme (Marie Skłodowska-Curie grant agreement No 799078) and by the Spanish MINECO and FEDER projects TIN2017-88213-R and UMA18-FEDERJA-003.

\bibliographystyle{IEEEtran}
\bibliography{IEEEabrv,biblio.bib}

\end{document}